\documentclass[aps,prl,showpacs,twocolumn]{revtex4}

\usepackage{graphicx}
\usepackage{amssymb}

\begin{document}

\title{
Coarse Nonlinear Dynamics and Metastability of Filling-Emptying Transitions:\\
Water in Carbon Nanotubes}
\author{Saravanapriyan Sriraman}
\email{rudram@princeton.edu}
\author{Ioannis G. Kevrekidis}
\email{yannis@princeton.edu} \affiliation{Department of Chemical
Engineering and PACM, Princeton University, Princeton, New Jersey 08544}
\break
\author{Gerhard Hummer}
\email{gerhard.hummer@nih.gov} \affiliation{Laboratory of Chemical
Physics, National Institute of Diabetes and Digestive and Kidney
Diseases, National Institutes of Health, Bethesda, MD 20892-0520}

\date{\today}

\begin{abstract}
  Using a Coarse-grained Molecular Dynamics (CMD) approach we study
  the apparent nonlinear dynamics of water molecules filling/emptying
  carbon nanotubes as a function of system parameters.
  Different levels of the pore hydrophobicity give rise to tubes that
  are empty, water-filled, or fluctuate between these two long-lived
  metastable states.
  The corresponding coarse-grained free energy surfaces and their
  hysteretic parameter dependence are explored  by linking MD to continuum fixed point
  and bifurcation algorithms. 
  The results are validated through equilibrium MD simulations.
\end{abstract}


\maketitle

\nobreak

Molecular dynamics (MD) simulations on classical or quantum energy
surfaces provide an atomically detailed description of complex
molecular processes such as protein folding or materials fracture.
However, such simulations are inherently limited by the requirement to
integrate accurately even the fastest molecular processes of bond
vibrations and atomic collisions.
Remarkable progress has been made recently in addressing the time scale
limitations of
MD (see, e.g., \cite{Voter:PRL:1997,Bolhuis:ARPC:2002,Schutte:JCompPhys:1999,Laio:PNAS:2002}).
To integrate out the fast molecular motions and explore the slow dynamics, we
are developing a ``coarse molecular dynamics'' (CMD) approach
\cite{HummerKevrekidis:JCP:2003,Kevrekidis:AICHE:2004}.
In principle, the projection operator formalism \cite{ZwanzigBook}
connects the microscopic dynamics to such slow motions.
However, exact analytical construction of the corresponding 
noise and memory terms is usually intractable.
In CMD, we circumvent this challenge by estimating 
\textit{on the fly} both the thermodynamic driving forces 
for slow motions and their dynamic properties.
Information about the slow coarse dynamics is extracted from the
projected motions of many, appropriately initialized, but otherwise
independent and unbiased replica simulations.

\begin{figure}[t]
{\includegraphics[width=0.75\columnwidth]{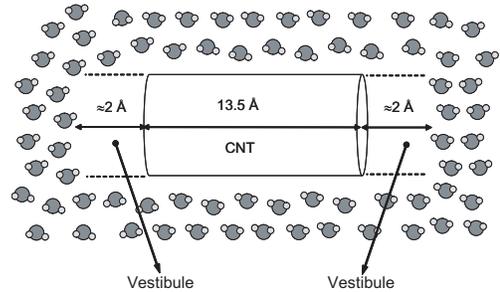}}
\caption{
  \label{vestib}
  Schematic of the CNT-water system showing the ``vestibules'' at the
  openings of the CNT and corresponding dimensions. The CNT is
  represented as a cylinder surrounded by water molecules.}
\end{figure}

In this letter, we extend the CMD approach to
explore directly {\it the physical parameter space} of a complex
molecular system.
This is accomplished by linking short bursts of MD simulations
  with continuum algorithms to simultaneously search phase \textit{and}
parameter space.
The general formalism is illustrated by studying the parametrically
controlled equilibrium between  the water-filled and
 empty state of a
short and narrow carbon nanotube (CNT) dissolved in water (Figure
\ref{vestib}).
The  CNT in water serves as a paradigm of a molecular system 
with nontrivial dynamic and thermodynamic behavior under parametric
control.
Conventional MD simulations \cite{Hummer:N:2001} showed that for
certain interaction strengths between the carbon atoms and water, the
molecular system exhibits ``bi-phasic'' character, with the tube
fluctuating on a nanosecond time-scale between a water-filled and an
empty state.
Such controlled fluctuations in the water occupancy have been
suggested as a mechanism to regulate water, proton, and ion transport
through biological protein pores
\cite{Hummer:N:2001,Beckstein:JPCB:01}. 
The computational prediction of filled tubes at strong
  water-tube interactions has since been confirmed experimentally for
  somewhat wider tubes
  \cite{Kolesnikov:PRL:2004}.
Important here is that, despite the molecular complexity, the
equilibrium thermodynamic calculations required to validate CMD are
feasible.
  
We perform classical MD simulations of a (6,6)-type nanotube of
  $\sim$13.5~{\AA} length and 8.1 {\AA} diameter in a box of
  $N_\mathrm{wat} = 1034$
  TIP3P water molecules \cite{Jorgensen:JCP:1983} under periodic
  boundary conditions with Ewald electrostatics and a time step of
  0.002 ps (additional simulation details as in Refs.
  \cite{Hummer:N:2001,Hummer:JPCB:2005}).
  We use CMD to explore the thermodynamics and kinetics of
  tube filling as a function of the ``hydrophobicity parameter''
  $\lambda$ scaling the attractive $r^{-6}$ carbon-water Lennard-Jones
  interactions: $u_\mathrm{C-O}(r) = A r^{-12} - \lambda C r^{-6}$
  where $A=696,790.7$ kcal mol$^{-1}$ {\AA}$^{12}$ and $C=564.5036$
  kcal mol$^{-1}$ {\AA}$^{6}$, respectively.
  In physical experiments, the effect of $\lambda$ on the
  water-affinity for the CNT pore can be mimicked by changing
  quantities like water pressure or osmolality.
To monitor filling and emptying, we use the water occupancy as a
coarse observation variable, $N=\sum_{i=1}^{N_\mathrm{wat}}
\Theta(r_{i},z_{i})$, where $r_{i}$ and $z_{i}$ are the water
radial and axial positions in the cylindrical coordinate system
defined by the instantaneous position and orientation of the freely
moving CNT (other simulation details as in \cite{Hummer:N:2001}).
The weight function is given by $\Theta(r,z) =
\exp[-{(r/R_{cyl})}^{6}-{(2z/L_{cyl})}^{6}]$ where
$L_{cyl}~=~17.5$~{\AA} and $R_{cyl}~=~4.05$~{\AA} are the length and
radius, respectively, of a cylindrical region that extends somewhat
beyond the $\sim$13.5~{\AA} long CNT to include water molecules near
the openings (see Figure  \ref{vestib}).
By including these ``vestibules'' of the CNT, we resolve the gradual
entry, or exit, of a water molecule without compromising the
information about the state of nanotube filling.
We found earlier that the tube appears bi-stable for $\lambda \approx 0.8$
\cite{Hummer:N:2001,Waghe:JCP:2002}; it is predominantly empty below
$\lambda \approx 0.7$, and filled above $\lambda \approx 0.9$.

Bi-stability can be succinctly summarized in
a {\it bifurcation diagram} that reports local maxima of the equilibrium
density  as a function of
  $\lambda$ and -- in the bi-stable  regime
-- also the intervening saddles.
In what follows, we will use short replica MD simulations to construct
 a kinetically-based coarse bifurcation diagram
and compare it to thermodynamics.  We will report the fixed points of
a timestepper constructed by CMD through the following steps.
In the ``lifting'' step,  we create
representative initial phase points of the full molecular system.
For a given value of the hydrophobicity parameter $\lambda$, we
harmonically bias the occupancy $N$ during a short, 15-ps MD run,
toward a target $N_0$ by adding $U_{bias}=k(N-N_0)^{2}/2$ to the
Hamiltonian ($k=50$ kcal mol$^{-1}$ for $N_0 \geq 1$;
$k=80$ kcal mol$^{-1}$ for $N_0 < 1$).
Five different conformations near each target $N_0$ value are saved during
the last 5 ps of the constrained run, and 10 sets each of random
Maxwell-Boltzmann velocities are assigned to create 50 starting
configurations.
In the ``evolve'' step, an \emph{unconstrained} MD run of $\tau =1$ ps 
duration is performed for each structure.
Finally, in the ``restrict'' step, the resulting trajectories are 
projected onto the coarse variable $N$.
In particular, the average occupancy $N$ at the end of the short
runs defines our CMD timestepper: $\overline{N(\tau;N_{0};\lambda)} \equiv
\Phi_{\tau}(N_{0};\lambda)$.

\begin{figure}[bt]
  {\includegraphics[width = 0.75\columnwidth]{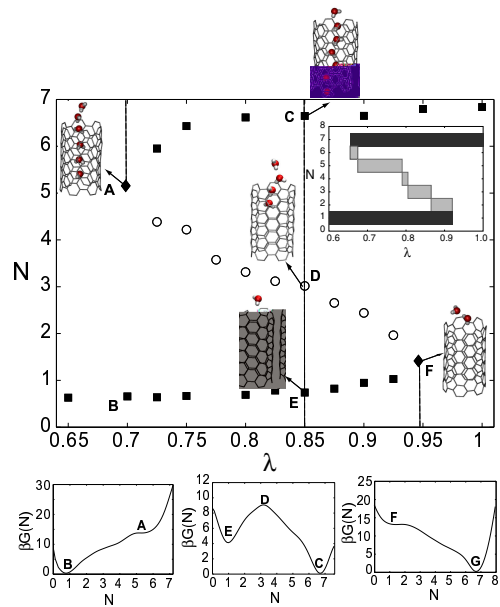}}
  \caption{
    \label{bifucn} Coarse bifurcation diagram for the CNT-water
    system. Solid filled squares ($\blacksquare$) correspond to fully
    filled ($N>5$) and empty
    states ($N\leq 1)$,
    respectively.
    Open circles ($\bigcirc$) correspond to partially filled states,
    while the filled diamonds ($\blacklozenge$) correspond to the
    turning points of the coarse bifurcation diagram.
    Vertical dashed lines indicate the upper (A) and lower (F)
    turning points, and $\lambda =
    0.85$ (points C-D-E), respectively.
    Representative structures for points A-F are indicated by
    arrows (carbon: gray; oxygen: red; hydrogen: white).
  The coarse effective free energy surfaces for
  $\lambda=0.7$, 0.85, and 0.95,
    respectively, are shown below the main figure.
    The inset in the top right corner of the main figure shows
      the effective bifurcation diagram obtained
      through histogram reweighting.
    }
\end{figure}

The fixed points for given values of $\lambda$ satisfy
\begin{equation}
  \label{Eqn_fixpt}
  f(N;\lambda) \equiv N - \Phi_{\tau}(N;\lambda) = 0.
\end{equation}
They correspond to metastable empty or full, as well
as to unstable, partially filled ``transition states''.
For small $\tau$ the above equation approximates the differential
(steady state) form ${\partial\Phi_t(N_{0};\lambda)}/{\partial t}
= 0$; all our coarse-grained computations can be equivalently
performed through this differential form.
Fixed points for given $\lambda$ are computed by solving
Eq.~(\ref{Eqn_fixpt}) through a Newton-Raphson type iteration with an
appropriate initial guess for $N$:

\begin{equation}
\label{Eqn_newtraph1} N_{1} = N_{0} -
\frac{f(N_{0};\lambda)}
{\partial f/ \partial N_{0}}
\end{equation}
In Eq.~(\ref{Eqn_newtraph1}), $\partial f /\partial N_{0}$ is
estimated by computing $N - \Phi_{\tau}(N;\lambda)$ for nearby initial
points $N$. 
Fixed points converged to within ${\cal O}(10^{-3})$.
The results  (fixed points as a function of $\lambda$) are plotted
in Figure \ref{bifucn}.

At the upper (lower) turning points we lose the filled (empty)
fixed points.
In principle, turning points can be found by computing the entire
diagram (using pseudo-arclength continuation \cite{Doedel});
  however, numerical bifurcation theory provides
more efficient algorithms that solve {\it directly} for these points
(and thus the boundaries of apparent hysteresis in parameter space).
In compact notation, $\mathbf{X}=[N,\lambda]^T$ is a vector describing
the position in the bifurcation diagram; the turning points are roots
of the vector-valued function $\mathbf{F}(\mathbf{X})= [ N -
\Phi_t(N;\lambda), 1-{\partial{\Phi_t(N;\lambda)}}/{\partial N}]^T =
\mathbf{0}$.
To solve these two coupled equations for the critical $N$ and
$\lambda$ simultaneously, we use a 2D recursive
Newton-Raphson iteration procedure:
\begin{equation}
  \label{Eqn_nrtp} \mathbf{X}_{1}=\mathbf{X}_{0} - {\Big[
    \frac{\partial \mathbf{F}}{\partial \mathbf{X}} \Big
    ]}_{\mathbf{X}_{0}}^{-1} \cdot \mathbf{F}(\mathbf{X}_0)
\end{equation}
The derivatives of the Newton-type iteration,
\begin{equation}
  \label{Eqn_jactp} \frac{\partial \mathbf{F}}{\partial \mathbf{X}}=
  \left[ \begin{array}{cc}
      1-\frac{\partial{\Phi_{\tau}(N;\lambda)}}{\partial N} &
      -\frac{\partial{\Phi_{\tau}(N;\lambda)}}{\partial \lambda} \\ 
      -\frac{\partial^{2}{\Phi_{\tau}(N;\lambda)}}{\partial N^{2}} &
      -\frac{\partial^{2}{\Phi_{\tau}(N;\lambda)}}{\partial N \partial
      \lambda}
    \end{array} \right]~,
\end{equation}
are estimated numerically from replica simulations initiated on a
centered ``stencil'' (i.e., grid) with $|\Delta N|=0.1$ and
$|\Delta\lambda|=0.025$ around $\mathbf{X}_0=[N_0,\lambda_0]^T$.
The ``lift-evolve-restrict'' procedure is implemented and the
timesteppers, $\Phi_{\tau}(N;\lambda)$, are computed for each node of
the grid.
Starting from a good initial guess, the procedure converges after a
few ($\sim$5) iteration steps: the updates for $[N,\lambda]^T$ and
residuals for $\mathbf{F}$ are of order ${\cal O}(10^{-3})$ and ${\cal
  O}(10^{-2})$, respectively.

Figure~\ref{bifucn} shows the resulting coarse bifurcation diagram for the
CNT-water system. 
The two metastable branches of the S-shaped diagram correspond to
empty ($N\approx 1$) and filled states ($N\approx 7$).
The lower and upper turning points are at $\lambda \approx 0.7$ and
0.95, respectively.

So far, we have taken an entirely dynamic perspective in our
computations.
To estimate the underlying thermodynamics, knowing that $N$ is a good
reaction coordinate \cite{Waghe:JCP:2002}, we assume diffusive
dynamics along $N$ in the form of an effective Fokker-Planck (FP) equation:
\begin{equation}
  \label{Eqn_FP} \frac{\partial P(N,t;\lambda)}{\partial t} = \big[
  -\frac{\partial}{\partial N} v(N;\lambda) +
  \frac{\partial^2}{\partial N^2} D(N;\lambda) \big] P(N,t;\lambda).
\end{equation}
The coarse timestepper calculations are precisely
what is needed to parametrize this FP:
\begin{equation}
  \label{Eqn_drift}
  v(N;\lambda) \equiv \frac{\partial
  \overline{N(t,N_{0};\lambda)}}{\partial t};~2D(N;\lambda) \equiv
  \frac{\partial \sigma^2[N(t;N_{0};\lambda)]}{\partial t}
\end{equation}
For short simulation times $\tau$, the coarse timestepper (the average
of the results of the replica simulations) is used to 
estimate the drift $v(N;\lambda)$; from the time-dependence of the
variance $\sigma^2[N(t;N_{0};\lambda)]$ of $N$, we estimate the diffusion
coefficient $D(N;\lambda)$.
In this notation, the effective free energy $G(N;\lambda)$ for the problem is
\begin{equation}
  \label{Eqn_effG} \beta G(N;\lambda) =
  -\int_{0}^{N}\frac{v(N';\lambda)}{D(N';\lambda)}dN' +
  \ln~D(N;\lambda) + \mathrm{const.}
\end{equation}
Maxima of the equilibrium density $\exp(-\beta G(N;\lambda))$ can be
found as zeroes of $v(N;\lambda) - D'(N;\lambda)$ [or, ignoring a weak
$N$ dependence of $D$, by the zeroes of $v(N;\lambda)$].
Clearly, the fixed points of our coarse timestepper are approximations
of the free energy extrema, where metastable and unstable
fixed points correspond to well bottoms and intervening saddles,
respectively.

In the inset of Figure~\ref{bifucn}, we show for
  comparison the
\textit{thermodynamic} bifurcation diagram, 
indicating the minima and saddles of the
equilibrium free energy surfaces $G(N;\lambda)$.
This diagram is estimated from three long (132 ns total) equilibrium
MD simulations, one in the predominantly
filled regime ($\lambda=1$), one in the predominantly empty regime
($\lambda\approx 0.75$), and one in between ($\lambda=0.785$).
By combining the 3-dimensional histograms of occupancy fluctuations
and water-CNT interaction energies using a weighted histogram method
\cite{Hummer:JPCB:2005,WHAM}, we  obtain the free energy
surface $G(N;\lambda)$ at different $\lambda$ values.
To avoid difficulties with insufficient sampling of the observation
variable in the equilibrium runs, we coarse-grained $N$ to the nearest
integer numbers.

Overall, we find excellent agreement between the coarse bifurcation diagram
from multiple short (1-ps) CMD runs, and that from long equilibrium
runs.
In particular, the location of the metastable and unstable branches is
nicely reproduced by the equilibrium results.
The main difference is that the lower turning point is at a somewhat
lower interaction strength $\lambda\approx 0.66$ compared to 0.7.
Such small differences can be due to several factors: (1) The
equilibrium-predicted lower turning point lies outside the
$\lambda$-range probed by MD and had to be extrapolated.
(2) The equilibrium computations were coarse-grained to integer $N$
values.
(3) The $N$-dependence of the effective diffusion coefficient was
neglected in constructing the kinetic bifurcation diagram; and finally
(4) CMD simulations at stationarity probe the dynamic stability of an
apparent fixed point \textit{over a given time horizon $\tau$.}
 The location of fixed as well as turning points from
CMD will depend somewhat on $\tau$. 
Here, the short runs are about 1 ps long, one order of magnitude shorter
than the duration of a typical water uptake or release event
$N\rightleftharpoons N+1$ \cite{Waghe:JCP:2002}, assuming
  Markov-chain dynamics.
For parameter values close to the turning points, as the free energy
wells become more shallow, typical transition times
$N\rightleftharpoons N+1$ become comparable to any fixed $\tau$.
As a result, one would expect the CMD turning points to move ``inwards,''
and the apparent hysteresis range to shrink 
\cite{Haataja:PRL:2004}.

\begin{figure}[t!h]
  {\includegraphics[width=0.75\columnwidth]{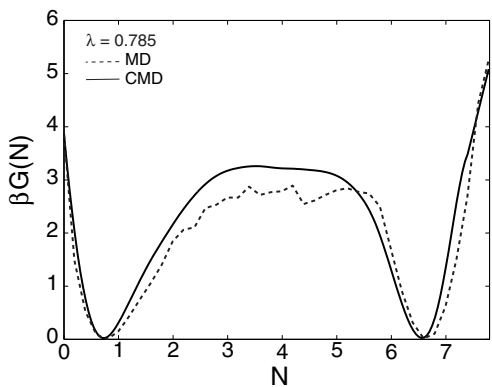}}
\caption{ \label{gvsncompare} Coarse free energy surface as a
  function of N for $\lambda=0.785$ where the empty and filled wells
  have equal depth. Results from equilibrium MD (dashed) and CMD [from
  Eq.~(\ref{Eqn_effG}); solid] are shown. The effective free energy
  predicted by CMD agrees well with that of 19-ns long equilibrium MD
  simulations.}
\end{figure}

Our CMD calculations have
so far been local in $N$-$\lambda$ space.  As an example of a
parametric search involving global properties, we next
locate the critical hydrophobicity parameter
value at which the filled and empty states have (approximately) equal
populations.
We formulate this as a fixed point problem that we solve iteratively.
Given an initial guess of  $\lambda$ we first solve
for the two metastable fixed 
points [$N_\mathrm{empty}(\lambda)$ and $N_\mathrm{full}(\lambda)$]. 
We then consider {\it any} reasonable path between the two
(parametrized by the coarse observable $N$), discretize it, and
estimate (through locally initialized and executed MD experiments and
quadrature) the state dependent drift and diffusion coefficient in the
FP equation, Eq.~(\ref{Eqn_FP}).
Finally, using Eq.~(\ref{Eqn_effG}), we compute
$G(N;\lambda)$.
To solve the equation $\label{deltaG} \Delta G
(\lambda) \equiv G(N_\mathrm{full};\lambda) -
G(N_\mathrm{empty};\lambda) = 0$  for the
  critical $\lambda$  we use a
contraction mapping with numerically estimated derivatives.
$G(N;\lambda)$ was computed as follows:
From different initial configurations for $0 \leq N_i \leq 8$, we
run 50 short ($\tau$ = 1 ps) replica simulations.
By robust linear fits to the evolution of
$\overline{N(t;N_{i};\lambda)}$ and $\sigma^2[N(t;N_{i};\lambda)]$,
we estimate $v(N)$ and $D(N)$ and compute $\beta G(N)$ from
Eq.~(\ref{Eqn_effG}) by integration through a smooth spline
approximation to the data.
As illustrated in Figure~\ref{gvsncompare}, the critical
$\lambda$ value is approximately 0.785, and this is
 confirmed by equilibrium MD simulations.
In particular, the MD and CMD free energy surfaces at this critical
parameter value are in excellent agreement.
We note that for equal populations in the two wells, the free-energy
gradients estimated by CMD are relatively small, resulting in
particularly reliable free energy surfaces, with larger deviations
expected in regimes where one or the other well dominates.
Finally, we note that CMD in combination with Kramers theory for
diffusive barrier crossing, produces accurate estimates of the rate
coefficients for escape from the metastable filled and empty states,
when compared with equilibrium MD [$\sim$1/(140 ps) vs. $\sim$1/(190
ps) from MD at the transition midpoint, $\lambda=0.785$].

We illustrated the  extension of the CMD approach to the exploration
of parameter space of
molecular systems.
Our computational methods are general and can be used to study
metastability boundaries in other systems (e.g., kinetic Monte Carlo
\cite{Haataja:PRL:2004,Makeev} or Brownian Dynamics \cite{Graham}).
Linking continuum nonlinear dynamics computation (fixed
point, continuation, integration or bifurcation algorithms) with
 MD
simulations has
the potential to accelerate parametric exploration of dynamic
as well as thermodynamic features.
The knowledge of good reaction coordinates is crucial for the success
of this approach; determining such coordinates based on data is the
subject of intense current research
\cite{Bolhuis:ARPC:2002,Hummer:JCP:2004,Nadler:2004}.

{\bf Acknowledgments} Support in part by the Intramural Research
Program of the NIH, NIDDK (GH) and NSF/ITR and DARPA/ARO (SS, IGK) is
gratefully acknowledged.


\end{document}